\documentclass[prl,showpacs]{revtex4}% Physical Review B
\usepackage{graphicx}% Include figure files
\usepackage{epstopdf}
\usepackage{bm}% bold math
\usepackage{natbib} % whatfor?
\usepackage{bbding} % whatfor?
\usepackage{placeins} %whatfor?
\usepackage{amsmath}
\usepackage{subfigure}
\usepackage{float}
\usepackage{hyphenat}

\usepackage[normalem]{ulem}

\newcommand{\Oh}{\mathrm{Oh}}
\newcommand{\fluct}[1]{\tilde{#1}}

\newcommand{\dynvisc}{\mu}     %dynamic viscosity
\newcommand{\density}{\rho}       %density
\newcommand{\surftens}{\sigma} %surface tension
\newcommand{\radius}{R}        %radius
\newcommand{\axpos}{x}         %axial position
\newcommand{\amppert}{\delta}
\newcommand{\relamppert}{\epsilon}
\newcommand{\length}{L}        %channel length
\newcommand{\vel}{u}           %velocity
\newcommand{\symbtime}{t}      %time
\newcommand{\growthrate}{\omega} %growth rate of a perturbation
\newcommand{\aspectratio}{\Gamma}

\newcommand{\dimlesswavenumber}{\kappa}

\newcommand{\removedtext}[1]{}

\begin{document}

\preprint{APS/123-QED}
\title{Stability of viscous long liquid filaments}% Force line breaks with \\

%\author{Theo Driessen} 
%\email{t.w.driessen@utwente.nl}
%\affiliation{University of Twente, Faculty of Science and Technology}
%
%\author{Roger Jeurissen}
%\affiliation{Eindhoven University of Technology, Department of Applied Physics}
%
%\author{Herman Wijshoff}
%\affiliation{Oce Technologies N.V., Venlo, The Netherlands.}
%
%\author{Federico Toschi}
%\affiliation{Eindhoven University of Technology, Department of Applied Physics}
%
%\author{Detlef Lohse} 
%\affiliation{University of Twente, Faculty of Science and Technology}

\author{Theo Driessen$^{a\ast}$\thanks{$^\ast$Electronic adress: t.w.driessen@utwente.nl}, Roger Jeurissen$^{b}$, Herman Wijshoff$^{c}$, Federico Toschi$^{b}$ and Detlef Lohse$^{a}$. \\\vspace{6pt} $^{a}${\em{Physics of Fluids group, Faculty of Science and Technology and Burgers Center of Fluid Dynamics, University of Twente, P.O. Box 217, 7500 AE Enschede, The Netherlands.}};
\\$^{b}${\em{Department of Applied Physics, Eindhoven University of Technology, PO Box 513, 5600 MB Eindhoven, The Netherlands.}}; \\$^{c}${\em{Oce Technologies N.V., Venlo, The Netherlands.}}}

\begin{abstract}

We study the collapse of an axisymmetric liquid filament both
analytically and by means of a numerical model. The liquid filament, also known as ligament, may either
collapse stably into a single droplet or break up into multiple
droplets. The dynamics of the filament are governed by the viscosity and the aspect ratio, and the initial perturbations of its surface.
We find that the instability of long viscous filaments can be completely explained by
the Rayleigh-Plateau instability, whereas a low viscous filament
can also break up due to end pinching.
We analytically derive the transition between stable collapse and breakup in the Ohnesorge number versus aspect ratio phase space.
Our result is confirmed by numerical simulations based on the slender jet approximation and explains recent experimental findings by Castr\'ejon-Pita
et al. , PRL \textbf{108}, 074506 (2012).
\end{abstract}

\maketitle
\section{Introduction}

The formation of liquid filaments is ubiquitous. It happens whenever there is droplet fragmentation \cite{Villermaux2007}.
Examples of fragmentation are the breakup of a liquid filament stretched from a bath \cite{Marmottant2004b}, or the collapse of a liquid film \cite{Bremond2005}.
The formation of these liquid filaments is very common in the breakup of ocean spume, where they influence the properties of the marine aerosols \cite{Veron2012}.
In industry, the dynamics of filaments are a key issue for the print quality in inkjet printing \cite{Wijshoff2010}, where elongated liquid filaments are ejected from the nozzle. 
For optimal print quality, the filaments should contract to a single droplet before they hit the paper. Therefore we study the details of the dynamics of these filaments.

Depending on the viscosity, size and shape of the filament, it may collapse into a single droplet, or it may break up into multiple smaller droplets \cite{Eggers2008}. A filament collapse resulting in a single droplet is called stable. A filament collapse into multiple droplets is called unstable.
The stability of the filament influences the number of droplets and their size distribution, after the collapse of the filament.

Since filaments are essentially finite liquid jets, the stability of liquid jets is relevant here. The earliest reported observations of the stability of liquid jets were done by Savart in 1833, who observed that a continuous liquid jet breaks up into distinct droplets \cite{Savart1833}.
Plateau reported that a varicose perturbation of the liquid jet is unstable, if the  wavelength of the perturbation is larger then the circumference of the jet \cite{Plateau1873}.
In 1878, Lord Rayleigh derived the dispersion relation for an infinite jet with a small periodical perturbation \cite{Rayleigh1878}.

The classical Rayleigh-Plateau instability assumes a sinusoidal perturbation with a time dependent amplitude
\begin{equation}
\radius(\axpos,\symbtime) = \radius_0+\fluct{\radius}(\symbtime)\cos\left(\dimlesswavenumber\frac{\axpos}{\radius_0}\right),
\end{equation}
where $\axpos$ is the axial position on the cylinder, and $\dimlesswavenumber$ is the dimensionless wavenumber of the varicose perturbation on the surface of the cylinder. 
The wavenumber is nondimensionalized by the relevant length scale of the problem, namely the unperturbed radius of the cylinder, $\radius_0$: $\dimlesswavenumber = 2\pi\radius_0\lambda^{-1}$. In this linear approximation, the amplitude of the perturbation $\fluct{\radius}(\symbtime)$ grows exponentially with time, 
\begin{equation}\label{eq:RP_growth}
\fluct{\radius}(\symbtime)=\amppert e^{\growthrate(\symbtime-\symbtime_0)},
\end{equation}
where $\growthrate$ is the growth rate of the perturbation amplitude, and $\amppert$ is the amplitude of the perturbation at $\symbtime = \symbtime_0$. In the inviscid theory by Rayleigh, $\growthrate$ is inversely proportional to the capillary time $\symbtime_{cap}=\sqrt{\frac{\rho\radius_0^3}{\surftens}}$, where $\rho$ and $\surftens$ are respectively the density and surface tension of the liquid.

Weber derived the disperion relation for the Rayleigh-Plateau instability including the influence of viscosity \cite{Weber1931}. The relevant dimensionless group for the dynamics of the viscous Rayleigh-Plateau instability on an infinite cylinder is the Ohnesorge number 
\begin{equation}\Oh_\radius = \dfrac{\dynvisc}{\sqrt{\density\surftens\radius_0}},\end{equation} 
where $\dynvisc$ is the dynamic viscosity. The Ohnesorge number is the ratio of the viscous timescale over the capillary time scale. The stability of the jet decreases with a decreasing Ohnesorge number. 
Weber showed in the same paper that the linearized dispersion relation of the Rayleigh-Plateau instability gives a very good approximation of the exact result. The linearized growth rate of the viscous Rayleigh-Plateau instability $\growthrate$ is \cite{Weber1931}
\begin{equation}\label{eq:growthrate}
\growthrate \symbtime_{cap} = \sqrt{\frac{1}{2}(\dimlesswavenumber^2-\dimlesswavenumber^4)+\frac{9}{4}\Oh_\radius^2\dimlesswavenumber^4} - \frac{3}{2}\Oh_\radius\dimlesswavenumber^2.
\end{equation}
For each value of $\Oh$, one dimensionless wave number has the highest growth rate: ${\dimlesswavenumber_{max}=(2+3\sqrt{2}\Oh_\radius)^{-1/2}}$. In this study we assume that the surface of the filament has a random varicose perturbation at $t=t_0$. A Rayleigh-Plateau instability originating from noise is dominated by $\dimlesswavenumber_{max}$. Substituting $\dimlesswavenumber_{max}$ into the dispersion relation, eq.(\ref{eq:growthrate}), we find that the growth rate of the fastest growing mode $\growthrate_{max}$ is a function of $\Oh_R$ only. When $\fluct{\radius}$ has grown to $\fluct{\radius} \approx \radius_0$, the filament breaks. Substituting $\fluct{\radius} = \radius_0$ and $\growthrate(\dimlesswavenumber,\Oh_R) = \growthrate_{max}(\Oh_R)$ into equation (\ref{eq:RP_growth}) gives the breakup time $\symbtime_B$,
\begin{equation}\label{eq:tbreak}
\symbtime_{B}=\symbtime_0+\dfrac{1}{\growthrate_{max}(\Oh_\radius)}\log\left(\dfrac{\radius_0}{\amppert}\right). 
\end{equation}
Experiments have shown that the linear approximation for the breakup time is a very good approximation \cite{Gonzalez2009}.

\begin{figure}[t]
\includegraphics[width=0.5\textwidth]{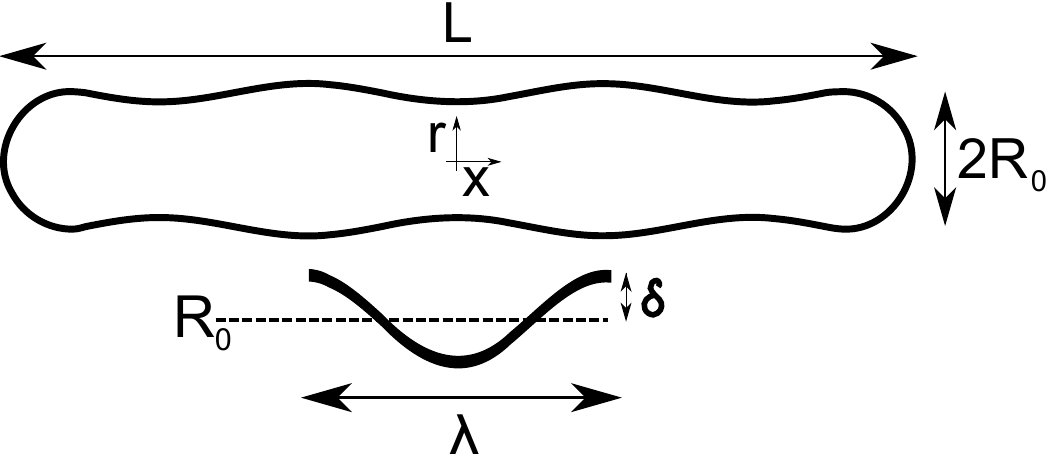}
\caption{Shape of the filament at $\symbtime_0$, just after the filament has pinched off from e.g. a film or a large filament. The surface is perturbed, with an assumed sinusoidal perturbation superposed on $\radius_0$. The wavelength of the perturbation is $\lambda$. The amplitude of the perturbation at $\symbtime_0$ is $\amppert$. The filament considered is symmetric in the $x=0$ plane. %The sinusoidal perturbation is essential, since the long viscous filamen etc.
}\label{fig:ic}
\end{figure}
In practice, liquid jets are never infinitely long. With the advance of computational power and experimental techniques, investigating the stability of liquid filaments became possible.
Using a numerical model, Schulkes found that above a certain $\Oh_\radius$, the filament collapse is always stable~\cite{Schulkes1996}. Later Notz \& Basaran refined this result by showing that the stability of the filament collapse depends on both $\Oh_\radius$ and the aspect ratio $\aspectratio=0.5\length/\radius$ of the filament~\cite{Notz2004}, where $\length$ is the length of the filament. So-called end pinching, first observed by Stone \cite{Stone1986}, is the cause of the breakup in both studies.

\begin{figure}[htp]
\includegraphics[width=0.5\textwidth]{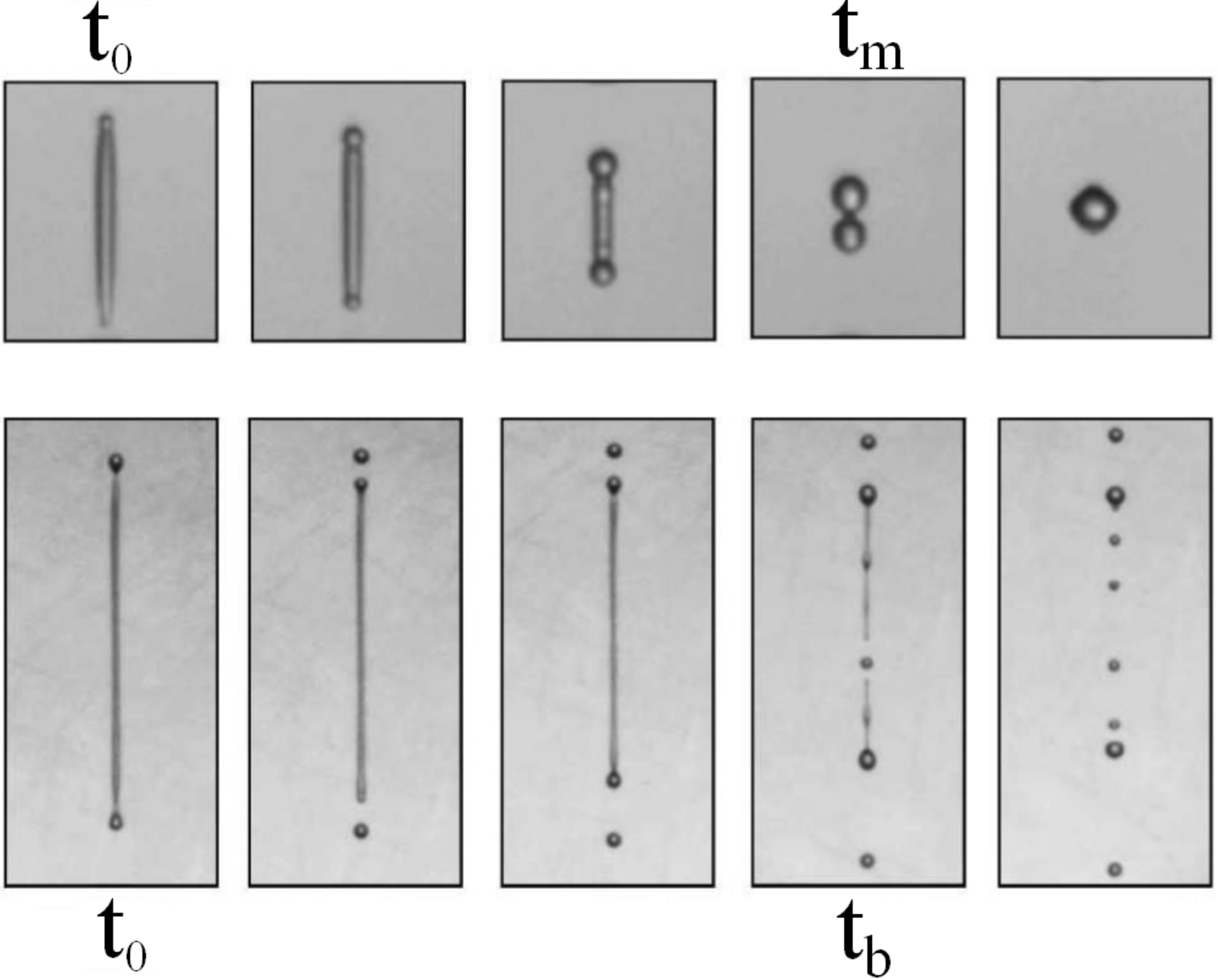}
\caption{Sequences of the contraction of a liquid filaments, taken from \cite{CastrejonPita2012} with permission. The upper sequence shows from left to right the stable collapse of a low viscous ligament. We call the moment when the tail droplets are in contact, $\symbtime_M$. The lower sequence shows the unstable collapse of a longer, more viscous ligament. At $\symbtime_0$ the filament breaks off from the big tail droplets. We call the moment when the filament breaks up everywhere $\symbtime_B$. The initial properties are $\aspectratio = 9$ and $\Oh_\radius =0.04$ for filament the upper filament, and $\aspectratio = 29.2$ and $\Oh_\radius =0.18$ for the lower filament.}\label{fig:example_filament}
\end{figure}

In a recent experimental study, Castr\'ejon-Pita \textit{et~al.}~\cite{CastrejonPita2012} validated the numerical results by Notz \& Basaran \cite{Notz2004} in the regime of $\Oh_\radius<0.1$, where end pinching is the dominant breakup mechanism.
The experiments however showed unstable filaments up to $\Oh_\radius \approx 1$. In the regime of long viscous filaments, the breakup occurs simultaneously along the entire axis of the jet, which is characteristic for the Rayleigh-Plateau instability on an infinite jet (figure \ref{fig:example_filament}).
In the existing numerical studies on the collapse of the finite liquid filament \cite{Schulkes1996,Notz2004}, the initial amplitude of the Rayleigh-Plateau instability has been set to zero. However, the breakup of the circular cylindrical part of the filament occurs within finite time, hence $\amppert$ must be nonzero at $\symbtime_0$. However, the experiments show that the circular cylindrical part of the filament breaks up. Hence $\amppert$ is finite at $\symbtime_0$.
The dynamics of the liquid filament are governed by three dimensionless groups, namely $\aspectratio$, $\Oh_\radius$ and the relative initial distortion $\relamppert = \amppert/\radius_0$.  

In this paper, the stability of long viscous filaments is predicted with the linear theory from \cite{Rayleigh1878,Weber1931}. The critical aspect ratio $\aspectratio_c$, above which the filament breaks up due to the Rayleigh-Plateau instability before it merges, depends on $\Oh_\radius$ and $\relamppert$. The results of the linear theory are validated against the experimental data \cite{CastrejonPita2012} and a drop formation model in the slender jet approximation \cite{Driessen2011} . For very high viscosity, an analogy is found with the viscosity dominated contraction of viscous sheets.

\section{Critical aspect ratio}
The critical aspect ratio $\aspectratio_c$ is defined as the aspect ratio above which the filament collapses due to the Rayleigh-Plateau instability. 
This aspect ratio is found by comparing the breakup time due to the Rayleigh-Plateau instability with the time that the filament needs to merge into a spherical droplet.
The analysis of the collapse of the liquid filament starts at $\symbtime_0$, when the filament pinches off from another body of liquid at both sides.
The filament contracts due to the surface tension.
As the filament contracts, bulges will form at the end of the filament.
These bulges are called tail droplets.
The fluid that has not yet been entrained by the tail droplets remains stationary.
For low to moderate viscosities, the typical velocity of these tail droplets can be found from balancing the momentum flux into the tail droplet with the capillary forces act on it.
This typical velocity is called the capillary velocity, $\vel_{cap}=\sqrt{\frac{\surftens}{\density\radius_0}}$ \cite{Schneider1966}.

When the collapse of the filament is stable, the tail droplets move towards the center of mass, where they merge into one droplet.
The distance that the tips of the jets have to travel is given by half the length of the filament minus the diameter of the tail droplet $2\radius_d$ at the moment of merging.
The minimal time for the tail droplets to merge is called the merge time $\symbtime_M$
\begin{equation}\label{eq:tmerge1}
\symbtime_{M}=\dfrac{0.5\length-2\radius_{d}}{\vel_{cap}},
\end{equation}
where $\radius_d$ is the radius of the tail droplets at the merge time $\symbtime_M$. At that time, each of the two tail droplets has accumulated half of the initial
volume of the liquid filament. This results in the tail
droplet radius
\begin{equation}\label{eq:radiusd}
\radius_d=\left(\frac{3}{4}\aspectratio\right)^{1/3}\radius_0 . 
\end{equation}
In the calculation of $\radius_d$ the filament is considered as a circular cylinder of length $L$. 
The merge time $\symbtime_M$ is rewritten as a function of $\aspectratio$ and $\symbtime_{cap}$, using equation (\ref{eq:tmerge1}), (\ref{eq:radiusd}), and $\vel_{cap}=\dfrac{\radius_0}{\symbtime_{cap}}$:
\begin{equation}\label{eq:tmerge}
\symbtime_{M}=\symbtime_{cap}\left[\aspectratio-\left(6\aspectratio\right)^{1/3}\right],
\end{equation}
this shows that $\symbtime_M$ is nearly proportional to $\aspectratio$, with a small correction since the tail droplets have a finite radius.

\begin{figure}[tp]
\includegraphics[width=.6\textwidth]{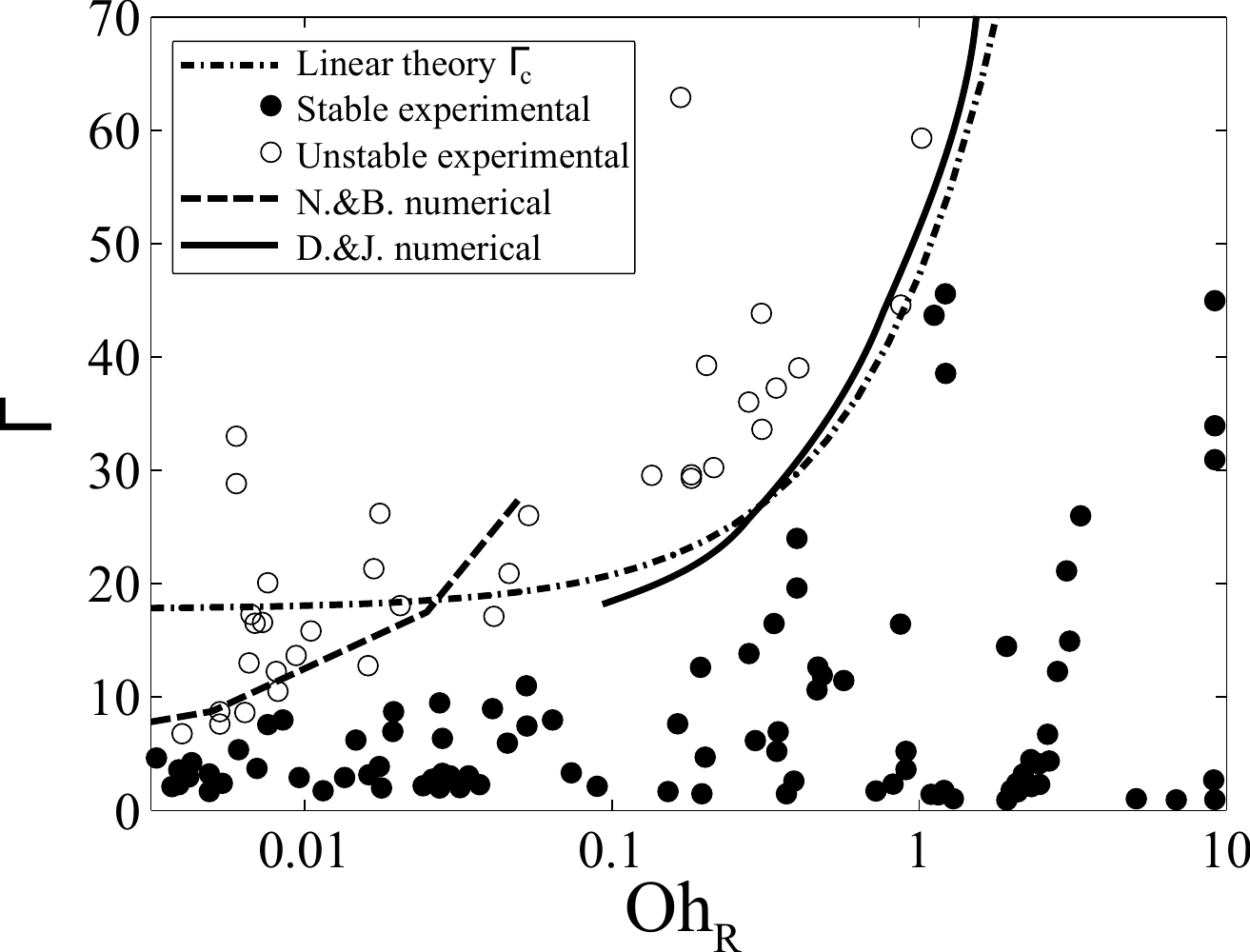}
\caption{Parameter study of the stability of the contraction of a filament. The result of the linear theory, $\aspectratio_c$, is shown for $\relamppert = 0.01$ in the dashed dotted line. The linear theory is confirmed by the results of the numerical model \cite{Driessen2011}, solid line. Stability of low viscous filaments has been studied numerically by Notz \& Basaran \cite{Notz2004}, dashed line. The experimental results by \cite{CastrejonPita2012} are shown (circles for unstable, disks for stable regime). Linear theory correctly predicts the critical aspect ratio for filaments of $Oh>0.1$.}\label{fig:phase_diagram}
\end{figure}

The Rayleigh-Plateau instability causes a filament to break up when $\symbtime_{B}<\symbtime_{M}$. When $\symbtime_{B}=\symbtime_{M}$, a slightly longer or less viscous filament would collapse into two or more droplets, rendering the collapse unstable. From this balance, the critical aspect ratio $\aspectratio_c$ can be found. An expression for the critical aspect ratio $\aspectratio_c$ as a function of $\growthrate_{max}(\Oh_\radius)$ and $\relamppert$ is found by balancing equations (\ref{eq:tbreak}) and (\ref{eq:tmerge}), putting $\symbtime_0=0$.
\begin{equation}\label{eq:Gammac}
 \frac{\log\left(\relamppert\right)}{\symbtime_{cap}\growthrate_{max}(\Oh_\radius)}+(6\aspectratio_c)^\frac{1}{3}-\aspectratio_{c} = 0 .
\end{equation}
In Fig. \ref{fig:phase_diagram} this critical aspect ratio is shown in the phase diagram of $\aspectratio$ versus $\Oh_\radius$, for $\relamppert = 0.01$.

When the viscosity dominates, the filament contraction is stable. Eggers \& Fontelos showed that isolated inertialess droplets cannot break up \cite{Eggers2005}. 
In the absence of inertia, the Rayleigh-Plateau instability no longer causes the filament to break up. Since the presented theory uses the Rayleigh-Plateau instability to find $\aspectratio_c$, it is of interest to find the upper viscous limit where the filament can still break up.
An analogy with the capillary contraction of a liquid filament is the capillary contraction of a liquid sheet  \cite{Taylor1959,Culick1960}. During the contraction, a liquid rim forms at the edge of the sheet, analogue to the tail droplet that forms in the contraction of a liquid jet. Brenner \& Gueyffier showed that the spatial extend of this rim can be estimated with the Stokes length, $\length_\dynvisc=\dynvisc^2/(\density\surftens)$ \cite{Brenner1999}. 
When $\length_\dynvisc$ is larger than the length of the sheet, the rim does not form and the sheet contracts homogeneously. We expect that the axial extend of the tail droplets on the viscous filaments is also the Stokes length. Viscosity dominates the contraction of a viscous filament when the Stokes length is larger than half the filament length, since the filament has two ends. The ratio between the Stokes length and half the length of the filament can be written as the length-based Ohnesorge number, $\Oh_\length$, with 
\begin{equation}
\Oh_\length^2=\frac{\dynvisc^2}{\density\surftens}\frac{2}{\length},\label{eq:ohl}
\end{equation}
where the first factor on the right-hand side is the Stokes length.
When $\Oh_\length\ll 1$, the Stokes length is much smaller than the length of the filament; this means that the influence of the axial extend of the tail droplet on the stability of the filament is negligible. The highest value of $\Oh_L$ in the unstable region in the phase diagram of Fig. \ref{fig:phase_diagram} is 0.03, which shows that the stabilizing effect found by \cite{Eggers2005} is not significant in this part of the phase space. 

\section{Numerical results}
We use a numerical model to validate the presented linear theory.
Equation (\ref{eq:tmerge}) shows that for large $\aspectratio$, the merging occurs after many $\symbtime_{cap}$, which is the timescale of the growth of perturbations. Even very small perturbations may cause a breakup before $\symbtime=\symbtime_M$ in the case of long filaments.
The visualization of the small initial perturbation demands a very high optical resolution. Dedicated experimental setups have been built to measure the initial perturbation amplitude of a stimulated Rayleigh-Plateau instability, such as \cite{Gonzalez2009, Bos2011}.
Since the Rayleigh-Plateau instability on the long viscous filaments grows from noise, the breakup events are different for all filaments.
As a result, this phenomenon cannot be observed stroboscopically.
In a numerical model on the other hand, the initial perturbation can be chosen with great precision. Therefore we use a numerical model to validate the linear theory.

The stability of an axisymmetric viscous liquid filament is modeled using the slender jet approximation \cite{Eggers1994,Shi1994,Wilkes1999,Notz2004,Hoeve2010,Driessen2011}. For this study we use the previously developed numerical model by Driessen \& Jeurissen \cite{Driessen2011}. In this model the singularities that occur at pinchoff and coalescence, are regularized below a radius that scales with the grid resolution; the influence of the regularization vanishes in the limit of vanishing spatial step size. For this research, this cut-off radius is set at $\radius_0/30$. 
The initial condition of the calculation is shown schematically in Fig. \ref{fig:ic}.
The grid is moving along with the center of mass of the filament.
The filament is considered to be axisymmetric and initially at rest. The perturbation is applied both on the radius and the axial velocity of the jet, using the linearized solution of the Rayleigh-Plateau instability \cite{Weber1931},
\begin{eqnarray}
\radius(0,\axpos)=\radius_0-\amppert\cos \left(\dimlesswavenumber_{max}\frac{\axpos}{\radius_0}\right ),\\
\vel(0,\axpos)=\frac{2\growthrate_{max}}{\dimlesswavenumber_{max}}\amppert\sin \left(\dimlesswavenumber_{max}\frac{\axpos}{\radius_0} \right).
\end{eqnarray}
The radial perturbation is symmetric in the origin, whereas the velocity fluctuation is anti-symmetric. The parameter scan is performed on the regime where the Rayleigh-Plateau instability is expected to dominate the breakup, $15<\aspectratio<70$ and $0.05<\Oh_\radius<2$. The value for $\relamppert$ is 0.01. This value has been chosen such that the result of the linear stability analysis matches with the stability results of the experimental data. Due to the experimental restrictions it is not possible to retrieve the values for $\relamppert$ for every separate experiment.
\begin{figure}[]
\includegraphics[width=\textwidth]{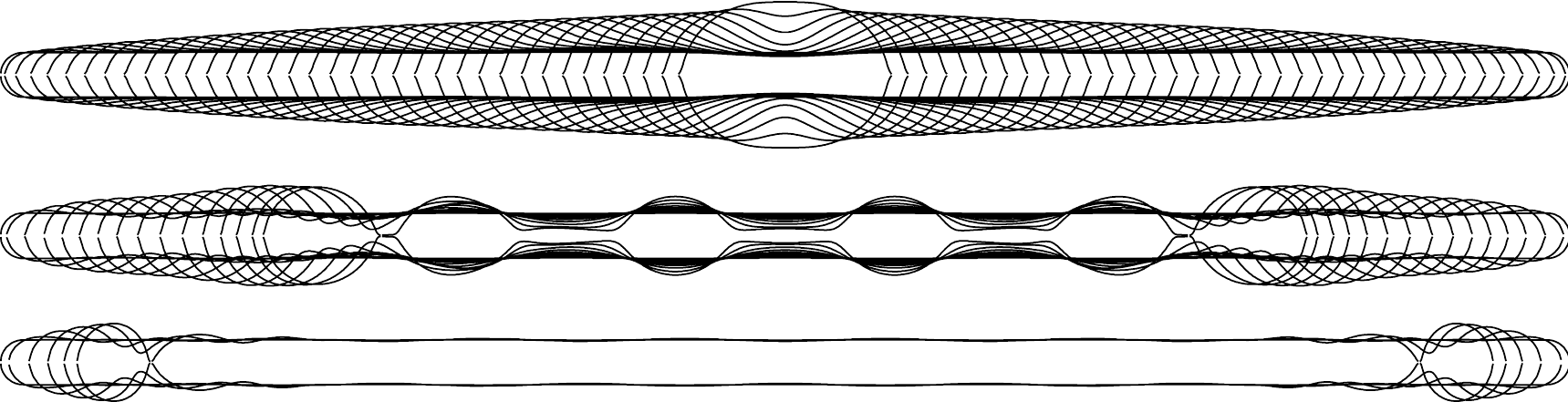}
\caption{Demonstration of the three different regimes, from top to bottom: Stable contraction, breakup due to the Rayleigh-Plateau instability, and breakup due to end pinching. For these numerical simulations, the Ohnesorge numbers are respectively 1, 0.1 and 0.01. The aspect ratio and the perturbation amplitude are kept constant at resp. $\aspectratio =35$ and $\relamppert=0.01$.
The time between the consecutive contours is $\symbtime_{cap}$.}\label{fig:numericalexamples}
\end{figure}
Each filament is simulated from $\symbtime_0$ to $\symbtime=\symbtime_M$. At the latter time, the tail droplets have merged in case of stable contraction. When the filament has broken up at any point in time before the end of the simulation, it is called unstable. The transition between the stable and unstable regime is shown in Fig. \ref{fig:phase_diagram}. With the results of the numerical model, it is shown that the stability of a viscous filament can indeed be predicted with linear theory, for a given $\relamppert$.

There are slight differences between the numerical results and the linear theory. Here we explain these deviations qualitatively.
At high viscosity, the numerical results are more stable than the linear theory. We hypothesize that this is due to the influence of the tail droplets on the liquid thread through viscosity. The linear theory assumes that the tail droplets are spherical when they merge. As can be seen in Fig. \ref{fig:numericalexamples}, this assumption is inaccurate for high viscous threads. The viscous dissipation that occurs in the tail droplet at low viscosity is smeared out over a larger region of the filament at high viscosity \cite{Brenner1999}. When these regions overlap, they stabilize the merging.
For low viscosity, the numerical model gives a less stable filament, since the capillary waves emitted by the tail droplets destabilize the merging \cite{Stone1986}. These capillary waves are not included in the linear theory. 

According to \cite{Schulkes1996, Notz2004, CastrejonPita2012}, one expects that the filament remains stable when $\Oh_\radius\geq1$, since end pinching does not occur in that region. This, however, does not mean that a filament with $\Oh_\radius \geq 1$ is intrinsically stable. A high viscous filament might break up due to the Rayleigh-Plateau instability, as we show here by the linear theory and the numerical model.

\section{Discussion} % Roger zegt na de conclusie, Federico voor de conclusie. Verder zegt federico: je moet geen paper willen schrijven over alle dingen die je niet gedaan hebt.
As the initial noise amplitude has not been measured directly, $\amppert$ can only be found indirectly from $\symbtime_B$. It is assumed to be the same for all experiments. To remove this limitation of the analysis, the noise amplitude must be studied with a dedicated setup.
% The stability of low viscous filaments is not discussed in this paper. The dynamics of the collapse of filaments in the low viscous regime can be quite complicated \cite{Notz2004}. Since the viscous damping is low, the energy goes to large amplitude surface oscillations resulting in exotic shapes.

At low viscosities, $\Oh_\radius\ll0.1$, the pinchoff resembles the shape of a double cone, in which one side bends back over the other with an internal angle larger than $90^o$ \cite{Day1998,CastrejonPita2012b}.
This geometry can not be described in the slender jet approximation, which requires the radius to be a single valued function of the axial position.
However, resolving this complicated geometry is not necessary to show that the stability of long viscous filaments is determined by $\aspectratio$, $\Oh_\radius$ and $\relamppert$, which was the objective of this paper.

\section{Conclusion}

A liquid filament either merges into a single droplet, or breaks up into multiple droplets. Breakup occurs if the aspect ratio $\aspectratio$ is larger than a critical value, which depends on the Ohnesorge number $\Oh_\radius$ and the relative perturbation amplitude $\relamppert$.
We provide a theoretical explanation of the filament stability in the regime of long viscous filaments, based on the Rayleigh-Plateau instability. 
At low viscosity, $\Oh_\radius< 0.1$, the critical aspect ratio is small due to end pinching. 
At higher viscosity, $\Oh_\radius> 0.1$, end pinching will not occur, but the filament may still break up due to the Rayleigh-Plateau instability. 
At very high viscosity, $\Oh_\length\gg1$, the filament collapse is stable. The transition to this regime is given by $\aspectratio_L=\Oh_R^2$.

\section{Acknowledgments}
We thank Jacco Snoeijer and Kundan Kumar for useful discussions. We are grateful for the valuable comments of I.M. Hutchings and his group, and the anonymous reviewers for their valuable comments on the manuscript. 
This work has been co-financed by the Dutch ministry of economical
affairs, Limburg Province, Overijssel Province,
Noord-Brabant Province, the partnership region Eindhoven
and by Oc\'e Technologies NV.

\end{document}